\pdfoutput=1

\documentclass{article}

     \usepackage[preprint, nonatbib]{neurips_2020}

\usepackage[utf8]{inputenc} 
\usepackage[T1]{fontenc}    
\usepackage{hyperref}       
\usepackage{url}            
\usepackage{booktabs}       
\usepackage{amsfonts}       
\usepackage{nicefrac}       
\usepackage{microtype}      

\usepackage[dvipsnames]{xcolor}
\usepackage{amsfonts}
\usepackage[export]{adjustbox}
\usepackage{textcomp}
\usepackage{gensymb}
\usepackage{graphicx}
\usepackage{hhline}
\usepackage{bbm}
\usepackage{multirow}
\usepackage{subcaption}
\usepackage{amsmath}
\usepackage[outdir=./Figures]{epstopdf}

\usepackage{arydshln}
\makeatletter
\newcommand{\printfnsymbol}[1]{%
  \textsuperscript{\@fnsymbol{#1}}%
}
\makeatother

\title{Graph Convolutional Networks Reveal Neural Connections Encoding Prosthetic Sensation}

\author{%
Vivek Subramanian\thanks{Equal contribution} \\
Duke University \\
vivek.subramanian@duke.edu
\And
Joshua Khani\printfnsymbol{1} \\
Duke University \\
joshua.khani@duke.edu
}

\begin{document}

\maketitle

\begin{abstract}

Extracting stimulus features from neuronal ensembles is of great interest to the development of neuroprosthetics that project sensory information directly to the brain via electrical stimulation. Machine learning strategies that optimize stimulation parameters as the subject learns to interpret the artificial input could improve device efficacy, increase prosthetic performance, ensure stability of evoked sensations, and improve power consumption by eliminating extraneous input. Recent advances extending deep learning techniques to non-Euclidean graph data provide a novel approach to interpreting neuronal spiking activity. For this study, we apply graph convolutional networks (GCNs) to infer the underlying functional relationship between neurons that are involved in the processing of artificial sensory information. Data was collected from a freely behaving rat using a four infrared (IR) sensor, ICMS-based neuroprosthesis to localize IR light sources. We use GCNs to predict the stimulation frequency across four stimulating channels in the prosthesis, which encode relative distance and directional information to an IR-emitting reward port. Our GCN model is able to achieve a peak performance of 73.5\% on a modified ordinal regression performance metric in a multiclass classification problem consisting of 7 classes, where chance is 14.3\%. Additionally, the inferred adjacency matrix provides a adequate representation of the underlying neural circuitry encoding the artificial sensation.

\end{abstract}

\section{Introduction}
A central objective in systems and computational neuroscience is characterizing the functional relationships that give rise to complex behaviors \cite{de2018connectivity,friston2011functional,bullmore2009complex}. The statistical correlation between activity in remote brain regions provides a framework for understanding the communication networks and organizational motifs that facilitate information processing in the brain \cite{schroter2017micro}. Additionally, it is of great interest to be able to classify neuronal events according to the preceding causes. For sensory neuroprosthetics, a primary goal is to predict what stimulation patterns lead to a specific neurophysiological outcome during a behavioral task \cite{de2018connectivity, panzeri2017cracking}. We present a graph convolutional network (GCN) model for predicting the stimulus features leading to the observed neurophysiological measurements. Additionally our model infers an adjacency matrix describing the effective connections underlying the integration of a new sensory modality facilitated through a sensory neuroprosthesis.

Recordings of cortical activity can be collected using implantable microarrays, which can sample the extracellular electrical potentials generated by populations of neurons. These time series can then be compared using a variety of methods to quantify how information is shared or transferred among the recorded populations. Understanding this connectivity can provide valuable insight into how the brain encodes external stimuli and can be used to explain differences in task performance from a mesoscopic level \cite{de2018connectivity}. Existing methods for computing functional connectivity are broadly divided into two groups: model-free and model-based \cite{de2018connectivity, friston2011functional}. Model-free methods -- including techniques such as correlation \cite{cohen2011measuring}, mutual information (MI) \cite{garofalo2009evaluation}, and transfer entropy ($i.e.$, Granger causality) \cite{orlandi2014transfer} -- are typically simpler and suffer from the fact that they cannot generate data for validation. On the other hand, model-based methods -- including generalized linear models \cite{song2013identification}, point process models \cite{lloyd2015variational, linderman2016bayesian}, and maximum entropy models \cite{roudi2015multi} -- can be computationally expensive to fit, and the nonlinearities they capture are limited to some functional form ($e.g.$, sigmoid or exponential). Both classes of methods fail to account directly for the effects of external stimuli.

GCNs are a class of deep-learning models used to extract latent representations from graphs \cite{kipf2016semi}, including those derived from social networks, molecular geometry, and traffic patterns \cite{wu2019comprehensive}. A single layer GCN is parameterized by an adjacency matrix $\mathbf{A} \in \mathbb{R}^{N \times N}$, where $N$ is the number of nodes in a graph, and a weight matrix $\mathbf{W} \in \mathbb{R}^{D \times D}$, where $D$ is the number of side features at each node. Each layer of a deep GCN nonlinearly aggregates features $\mathbf{X} \in \mathbb{R}^{N\times D}$ from neighboring nodes to produce hidden representations as:
\begin{align}
    \mathbf{H}(\mathbf{X}; \mathbf{A}, \mathbf{W}) = f(\mathbf{A}\mathbf{X}\mathbf{W})
    \label{eqn:gcn}
\end{align}
where $f$ is a nonlinear activation function. Rows of $\mathbf{A}$ denote edges through which information can flow in a graph: two nodes indexed by $i$ and $j$ have an edge between them if $A_{ij}\neq 0$. Edges can be binary or weighted depending on the application.

While GCNs have shown promise in a wide range of regression and classification tasks \cite{wu2019comprehensive}, it is standard to specify a known adjacency matrix when fitting a GCN. For instance, when trying to predict whether two proteins will interact, one can specify an adjacency matrix describing whether pairs of amino acids are connected \cite{fout2017protein}. When quantifying effective connectivity between populations of neurons, this matrix is precisely what we wish to infer.

We propose a GCN-based method to extract stimulus-predictive features from multi-unit spiking activity. Other baseline models simply treat the spike count correlation as a measure of effective connectivity. Our model, however, infers an adjacency matrix that captures similar correlation while also containing unique structure that is optimal for predicting the IR light stimulus; these networks are likely to represent the networks involved in encoding the prosthetic sense. In the future, a transform function used in a sensory neuroprostheis could be implemented that aims to elicit a specific neuronal response in order to improve behavioral performance and enhance embodiment of the exogenous sensory modality.

\section{Data Collection and Preprocessing}

\begin{figure}
    \centering
    \includegraphics[width=.6\textwidth]{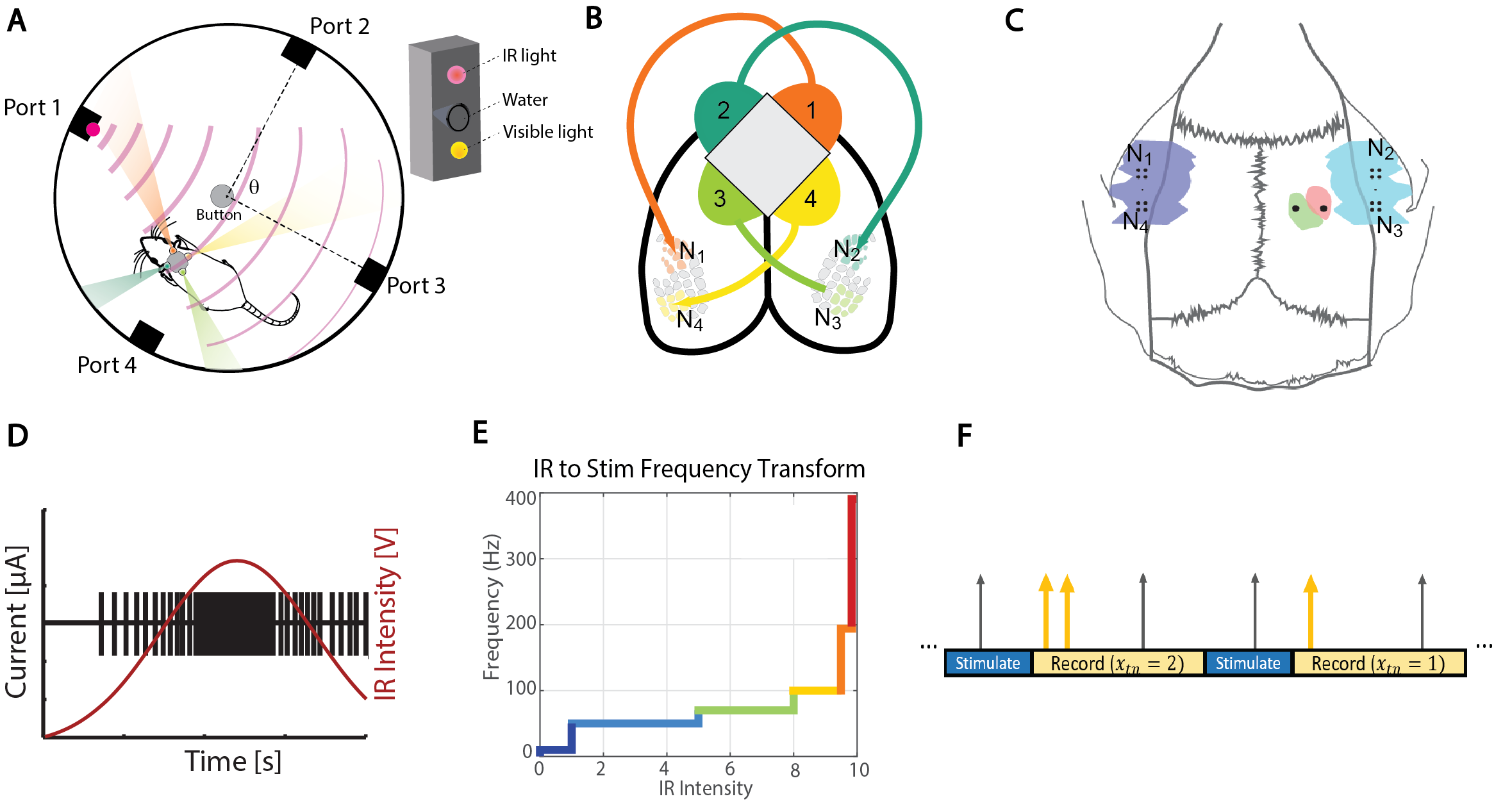}
    \caption{Experimental design and trial structure. A) Schematic of behavioral chamber used for the IR navigation task with four stimulus-reward ports along the perimeter. (\textit{top-right}) Each port consists of visual LED, IR LED, and conical recess with a photobeam across the opening and water spout. B) Schematic of IR sensors (1-4) along headcap to the corresponding stimulation sites (N$_1$-N$_4$). C) Dorsal view of rat skull with overlay of functional brain regions (S1L, dark blue; S1R, light blue; POm, green; VPM, red). Black dots indicate electrode pairs (S1) and cannulated bundles (VPM and POM). D) Stimulation pulse frequency increases with higher IR intensity. E) IR-intensity-to-frequency transform function follows a step-wise exponential function. Intensity of each sensor transformed independently of the rest. F) Stimulation and recording period for each stimulation event. During each recording period the number of spikes occurring during the first 70 ms are counted and treated as regressors for the model.}
    \label{fig:exprDesign}
\end{figure}

We use data from a single rat performing a sensory-augmented navigation task using a chronically-implanted sensory neuroprosthesis which projects information about the direction and distance to an infrared (IR) light source onto the animal's primary somatosensory (S1) cortex \cite{hartmann2016embedding,thomson2017cortical}. All behavioral tasks and surgical procedures were conducted in accordance with the National Research Council’s Guide for the Care and Use of Laboratory Animals and were approved by the Duke University Institutional Animal Care and Use Committee (IACUC). Recordings of neural activity are acquired in S1 as well as two thalamic nuclei, the ventroposteriomedial (VPM) and posteriomedial (POM) nuclei, which exchange dense  bidirectional projections with the stimulated region of S1 \cite{temereanca2004functional}. Further details on the behavioral set up and the IR-sensing prosthetic system can be found elsewhere \cite{thomson2013perceiving, hartmann2016embedding,thomson2017cortical}.

Briefly, behavioral experiments were conducted in a cylindrical chamber, which the animal was able to freely navigate (Figure \ref{fig:exprDesign}A). Stimulus-reward ports along the circumference consisted of an IR LED and a water reward spout. The animal initiated trials by breaking a photobeam in the central button, which activated the IR LED bulb at a randomly selected port. Selection of the activated port was rewarded with a variable amount of water. Multi-electrode microarrays, consisting of stimulating and recording electrodes, were implanted bilaterally in S1 (Figure \ref{fig:exprDesign}B). A magnetically mounted, detachable headcap houses four IR sensors each coupled to a specific stimulation site in S1, selected to be topographically "natural" when compared to innate, contralateral sensory processing paradigms (e.g. front, left sensor coupled to right, anterior stimulation site). In addition to the recording electrodes in S1, cannulated bundles of recording electrodes were implanted in VPM and POM (Figure \ref{fig:exprDesign}C).

Neural recordings were collected over 7 behavioral sessions consisting of between 113 to 201 trials (Table \ref{tab:mainresults}). A step-wise exponential IR-intensity-to-frequency transform function was implemented (Figure \ref{fig:exprDesign}D-E). When an IR sensor detected IR light from the activated LED, stimulation pulse trains were delivered to the corresponding S1 site. Within each trial, the stimulation pulse trains were divided into stimulating (70ms) and recording (140ms) periods (Figure \ref{fig:exprDesign}F). Regressors for our model consisted of spike counts detected during the first half of the recording period following the cessation of stimulation for each stimulation event. Stimulation artifact was removed in post-processing. The final 42 recording channels used in our model were those that contained clear neuronal activity across all 7 sessions. This allowed us to compare adjacency matrices inferred for each session across the individual sessions. Throughout this paper, we use the terms neurons, units, and electrodes, interchangeably. Since we solely focus on multi-unit activity, spiking activity detected by each individual recording electrode represents one unit, or node, in our analysis.

\section{Model}
Our goal is to predict the stimulus frequency transduced from each sensor while simultaneously inferring the effective connectivity between the neuronal populations. For simplicity, these frequencies $f \in \{0, 10, 20, 50, 100, 200, 400\}$ Hz are mapped to integers $y_{ts} \in \{0, 1, 2, 3, 4, 5, 6\}$ (notation explained below). We consider two neural network architectures to accomplish this goal: one which explicitly takes into account previous spiking history and another which captures this through a time-evolving hidden state.

\subsection{Spike history GCN (SH-GCN)}
\begin{figure}[!ht]
    \centering
    \includegraphics[width=0.35\textwidth]{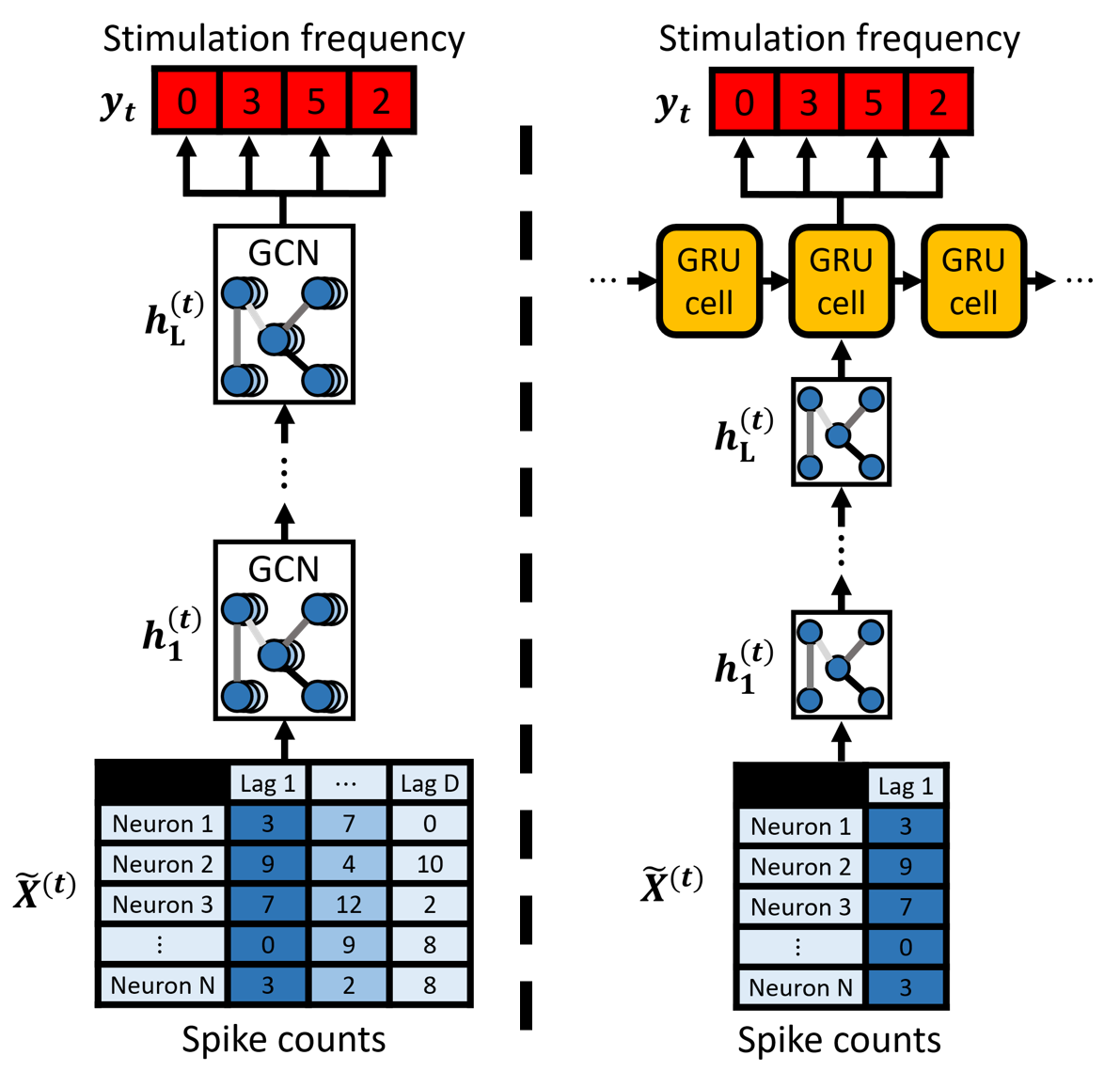}
    \caption{Two model architectures. The spike-history GCN (SH-GCN, left) explicitly accounts for past information by treating previous spike counts $\tilde{\mathbf{X}}^{(t)}_{nd}$ for $d = 1, \ldots, D$ as GCN node side features. The recurrent GCN (GRU-GCN, right) maintains a hidden state via a GRU cell that is updated with new samples. In both cases, the inferred GCN adjacency matrix is shared across hidden layers.}
    \label{fig:models}
\end{figure}
Figure \ref{fig:models} (left) shows the architecture of the SH-GCN model. The inputs and outputs are given by $\mathbf{X} \in \mathbb{R}^{T \times N}$ and $\mathbf{Y} \in \{0, \ldots, 6\}^{T \times |\mathcal{S}|}$, where element $x_{tn}$ of $\mathbf{X}$ represents the number of spikes generated by neuron $n = 1, \ldots, N$ during time bin $t = 1, \ldots, T$ and element $y_{ts}$ of $\mathbf{Y}$ represents the corresponding light levels measured on IR sensors $s \in \mathcal{S} = \{1, 2, 3, 4\}$. The input layer is a GCN with $N$ nodes containing $D$ features: each node corresponds to a single neuron and the $D$ features represent the history of spike counts for that neuron. To derive $D$ spike history features, for each time bin $t$, we concatenate rows $t-D+1:t$ of $\mathbf{X}$ along a new axis to create a new tensor $\tilde{\mathbf{X}} \in \mathbb{R}^{(T-D+1) \times D \times N}$. For each time point, a feature matrix $\tilde{\mathbf{X}}^{(t)} \in \mathbb{R}^{D \times N}$ is input into the GCN.

Graph convolution operations given by (\ref{eqn:gcn}) are applied sequentially to the input and hidden representations of each $\tilde{\mathbf{X}}^{(t)}$, yielding $\tilde{\mathbf{H}}^{(t)}_l$, where $l = 1, \ldots, L$ indexes the number of hidden GCN layers. Both the adjacency matrix $\mathbf{A} \in \mathbb{R}^{N \times N}$ and weight matrices $\mathbf{W}_l \in \mathbb{R}^{D\times D}$ are square; hence, the number of nodes and number of side features is preserved across GCN layers. Moreover, like a standard GCN, the adjacency matrix is shared across GCN layers while the weight matrix is unique to each layer. Intuitively, the adjacency matrix is a measure of connectivity and allows information to flow among neurons while the weight matrix allows features of individual neurons to mix to create meaningful, abstract representations. Each node represents a single neuron, and the number of hops away from which a node aggregates information scales linearly with the number of GCN layers.

The adjacency matrix is parameterized as:
\begin{align}
    \mathbf{A} = \tanh{\mathbf{Z}}
\end{align}
where $\tanh$ is applied elementwise and $\mathbf{Z} \in \mathbb{R}^{N\times N}$ is itself parameterized as $\mathbf{Z} = \tilde{\mathbf{A}}\tilde{\mathbf{A}}^\top$, where $\tilde{\mathbf{A}} \in \mathbb{R}^{N\times P}$. The resultant symmetric adjacency matrix can be interpreted as a nonlinear analog of correlation which explains connectivity between recorded neurons. We chose $2 = P << N = 42$ by performing principal component analysis (PCA) on the time series. In all sessions, $P = 2$ PCs were enough to represent 60\% of the variance in the data. Hence, we restricted the rank of $\mathbf{A}$ to be at most two, allowing the remaining model parameters to capture the rest of the variance.

\subsection{Recurrent GCN (GRU-GCN)}
Figure \ref{fig:models} (right) shows the architecture of the recurrent GCN (GRU-GCN) model. The model is closely related to the baseline model in that it also utilizes hidden GCN layers. However, instead of treating spike history as node features, we employ a gated recurrent unit (GRU) to keep track of a hidden state \cite{cho2014learning}. This hidden state is advantageous for two primary reasons: (1) it implicitly allows the model to maintain information from the previous time point all the way back to the beginning of the time series and (2) the state vector integrates both information from spike counts and sensor levels as it trained using teacher forcing. On the other hand, the GRU-based model is more highly parameterized. Specifically, the number of parameters in a GRU is given by $3(n^2 + nm + n)$ where $n$ is the number of output units and $m$ is the number of input units. We will discuss this tradeoff in further detail below.

\subsection{Ordinal regression loss function}
As described above, the stimulation frequency is discretized into seven levels. Depending on the intensity of the light detected by the IR light sensors, stimulation at one of those frequencies is delivered to S1. While this can be thought of as a multi-output, multiclass classification problem for which we can employ a softmax cross-entropy loss function between the logits generated by the network and one-hot vector encodings corresponding to each light level, this cost function does not penalize the weights in proportion to how grossly misclassified the predictions are.

Instead, we employ an ordinal regression (OR) loss function. Intuitively, OR lies between classification and regression. Classes are ordered, so the greater the deviation from the true class label, the greater the penalty. A common mathematical framework for OR involves estimating the probability of a continuous latent variable $y^*$ lying between two thresholds $\theta_{i-1}$ and $\theta_{i}$. This latent variable $y^*_t$ may come from any generative process (e.g., a neural network $g_{\mathbf{W}}(\mathbf{x})$ with parameters $\mathbf{W}$) and is typically assumed to follow a standard normal distribution, conditional on the input $x_t$. Formally, 
\begin{align}
    P(y = i) &= P(\theta_{i-1} < y^* \leq \theta_{i}) \nonumber\\
    &= P(\theta_{i-1} < g_\mathbf{W}(\mathbf{x}_t) + \epsilon_t \leq \theta_{i}) \nonumber\\
    &= P(\theta_{i-1} - g_\mathbf{W}(\mathbf{x}_t) < \epsilon_t \leq \theta_{i} - g_\mathbf{W}(\mathbf{x}_t)) \nonumber\\
    &= \Phi(\theta_{i} - g_\mathbf{W}(\mathbf{x}_t)) - \Phi(\theta_{i-1} - g_\mathbf{W}(\mathbf{x}_t))
\end{align}

The thresholds $\theta_i$ for  $i = 1, \ldots, K-1$, where $K$ is the number of classes, are nondecreasing and must be learned along with the model parameters $\mathbf{W}$. The $P(y = i)$ can then be interpreted as the parameters of a categorical distribution, so minimizing the negative log likelihood will result in maximum likelihood estimates for $\mathbf{W}$ and $\boldsymbol{\theta} = [\theta_1, \cdots, \theta_{K-1}]$.
While this function is fully differentiable, in practice, we found that implementing it directly to perform OR it results in slow convergence and sensitivity to initialization.

A popular method performing OR with neural networks was proposed by \cite{cheng2008neural}. The authors reduce the OR problem to a set of binary classification problems. Each classifier is associated with a learned threshold, and the the predicted rank is given by the sum of the outputs which exceed a chosen cutoff. This method was also employed by \cite{niu2016ordinal} to determine the age of a subject depicted in an image using a convolutional neural network and by \cite{fu2018deep} to estimate depth from single images (where depth is discretized into intervals whose boundaries must be estimated).

Formally, the class labels for each output $s = 1, 2, 3, 4$ are encoded by a $(K-1)$-hot vector, with class $k \in \{0, \ldots, K-1\}$ represented by a vector with the first $k$ elements set to $1$. For each sample $t = 1, \ldots, T$, we perform $K-1$ binary classifications for each output using a binary cross entropy loss function, indexed by $i = 0, \ldots, K-2$:
\begin{align}
    \mathcal{L}^{(t)}_{si} &= -y^{(t)}_{si}\log(\hat{y}^{(t)}_{si}) - (1 - y^{(t)}_{si})\log(1 - \hat{y}^{(t)}_{si})
    \label{eqn:loss}
\end{align}
where $\hat{y}^{(t)}_{si} = \sigma(g_s(\tilde{\mathbf{X}}^{(t)}) - \theta_{si})$, $g_s(\tilde{\mathbf{X}}^{(t)})$ is a scalar-valued prediction made by our model for output $s$, and $\sigma(\cdot)$ is the sigmoid activation function. Losses are accumulated over time and used to train the network weights $\mathbf{W}$ and the thresholds $\boldsymbol{\Theta} = [\boldsymbol{\theta}_1, \ldots, \boldsymbol{\theta}_{|S|}]$ via backpropagation. We choose to use this method of ordinal regression because of its practical efficacy and straightforward implementation.

\subsection{Training specifications}
We fit each of our models using the Adam optimizer with a learning rate of 0.001, $\beta_1 = 0.9$, $\beta_2 = 0.999$, and $\epsilon = 10^{-8}$, the default values in TensorFlow. The performance was consistent across a wide range of optimization hyperparameters. Models were trained on an NVIDIA Titan XP GPU for up to 3000 steps or until average training loss over the past five steps was found to decrease by less than $0.00005$\%.

\section{Results}
\subsection{Evaluation metrics}
We compare our models on an OR performance metric modified from  \cite{cheng2008neural, niu2016ordinal}, and defined as:
\begin{align}
    \bar{F}_{OR} = 1 - \sqrt{\frac{1}{T|\mathcal{S}|(K-1)}\sum_{t, s, k} \left(y_{sk}^{(t)} - \mathbbm{1}[\hat{y}_{sk}^{(t)}]\right)^2}
\end{align}
where $\mathbbm{1}[z] = 1$ if $z > 0.5$ and $0$ otherwise. $\bar{F}_{OR}$ falls between 0 and 1 with larger values indicating better performance. Like (\ref{eqn:loss}), this penalizes for class labels that are far in rank from the true label. For reference, we also report average, weighted $F_1$ score defined as:
\begin{align}
    \bar{F}_1 = \frac{1}{|\mathcal{S}|} \sum_{s,i} 2\frac{p_{si}\cdot r_{si}}{p_{si} + r_{si}} n_i
\end{align}
where $n_i$ is the fraction of points belonging to class $i = 0, 1, \ldots, K-1$ and $p_{si}$ and $r_{si}$ are, respectively, the precision and recall for the $i$\textsuperscript{th} label on the $s$\textsuperscript{th} output.

\begin{table*}[!ht]
\caption{Out-of-sample performance of baseline and proposed models on seven sessions.}
\begin{center}
\adjustbox{max width=\textwidth}{
\begin{tabular}{c|c|c|c|c|c|c|c|c}
Session & Trials & Model & Loss function & $L$ & $D$ & \# params & $\bar{F}_1$ & $\bar{F}_{OR}$ \\
 & (Samples) & & & & & & & \\ \hhline{=|=|=|=|=|=|=|=|=}
 &  &  GRU-GCN & Ordinal & 3 & 1 & 1425 & 0.578 & \textbf{0.652} \\
 &  &  GRU & Ordinal & 0 & 1 & 1296 & 0.567 & 0.649 \\ \cdashline{3-9}
1 & 113 & SH-GCN & Ordinal & 5 & 1 & 579 & 0.577 (0.456) & 0.649 (0.572) \\
 & ($17.0 \pm 7.4$) &  SH-GCN & Cross-entropy & 3 & 1 & 1501 & 0.558 & \textbf{0.651} \\
 &  &  Feedforward & Cross-entropy & 3 & 1 & 1353 & 0.536 & 0.633 \\ \hline
 &  &  GRU-GCN & Ordinal & 3 & 1 & 1425 & 0.642 & \textbf{0.697} \\
 &  &  GRU & Ordinal & 0 & 1 & 1296 & 0.622 & 0.676 \\ \cdashline{3-9}
2 & 201 & SH-GCN & Ordinal & 3 & 5 & 1741 & 0.619 (0.603) & \textbf{0.678} (0.670)\\
 &  ($10.4 \pm 4.3$)  &  SH-GCN & Cross-entropy & 3 & 1 & 1501 & 0.574 & 0.672 \\
 &  &  Feedforward & Cross-entropy & 3 & 1 & 1353 & 0.566 & 0.649 \\ \hline
 &  &  GRU-GCN & Ordinal & 1 & 1 & 1425 & 0.651 & \textbf{0.689} \\
 &  &  GRU & Ordinal & 0 & 1 & 1296 & 0.638 & 0.688 \\ \cdashline{3-9}
3 & 145 & SH-GCN & Ordinal & 3 & 3 & 1105 & 0.614 (0.508) & 0.678 (0.626)\\
 & ($14.6 \pm 5.7$) &  SH-GCN & Cross-entropy & 3 & 1 & 1501 & 0.623 & 0.673 \\ \textbf
 &  &  Feedforward & Cross-entropy & 3 & 1 & 1353 & 0.641 & \textbf{0.696} \\ \hline
 &  &  GRU-GCN & Ordinal & 1 & 1 & 1425 & 0.699 & \textbf{0.735} \\
 &  &  GRU & Ordinal & 0 & 1 & 1296 & 0.657 & 0.699 \\ \cdashline{3-9}
4 & 169 & SH-GCN & Ordinal & 3 & 5 & 1741 & 0.678 (0.683) & \textbf{0.720} (0.718) \\
 & ($15.2 \pm 4.7$) &  SH-GCN & Cross-entropy & 3 & 1 & 1501 & 0.635 & 0.702 \\
 &  &  Feedforward & Cross-entropy & 3 & 1 & 1353 & 0.681 & 0.712 \\ \hline
 &  &  GRU-GCN & Ordinal & 1 & 1 & 1296 & 0.604 & \textbf{0.683} \\
 &  &  GRU & Ordinal & 0 & 1 & 1425 & 0.595 & 0.674 \\ \cdashline{3-9}
5 & 134 & SH-GCN & Ordinal & 3 & 3 & 1105 & 0.598 (0.499) & \textbf{0.676} (0.636) \\
 & ($16.4 \pm 5.1$) &  SH-GCN & Cross-entropy & 3 & 1 & 1501 & 0.563 & 0.670 \\
 &  &  Feedforward & Cross-entropy & 3 & 1 & 1353 & 0.595 & 0.674 \\ \hline
 &  &  GRU-GCN & Ordinal & 1 & 1 & 1425 & 0.617 & 0.711 \\
 &  &  GRU & Ordinal & 0 & 1 & 1296 & 0.648 & \textbf{0.718} \\ \cdashline{3-9}
6 & 152 &  SH-GCN & Ordinal & 3 & 3 & 1105 & 0.590 (0.542) & 0.706 (0.697)\\
 & ($15.4 \pm 4.9$) & SH-GCN & Cross-entropy & 3 & 1 & 1501 & 0.565 & 0.698 \\
 &  &  Feedforward & Cross-entropy & 3 & 1 & 1353 & 0.624 & \textbf{0.718} \\  \hline
 &  &  GRU-GCN & Ordinal & 1 & 1 & 1425 & 0.621 & \textbf{0.678} \\
 &  &  GRU & Ordinal & 0 & 1 & 1296 & 0.602 & 0.671 \\ \cdashline{3-9}
7 & 120 & SH-GCN & Ordinal & 3 & 1 & 493 & 0.557 (0.585) & \textbf{0.637} (0.622) \\
 & ($16.2 \pm 5.0$) & SH-GCN & Cross-entropy & 3 & 1 & 1501 & 0.560 & 0.632 \\
 &  & Feedforward & Cross-entropy & 3 & 1 & 1353 & 0.531 & 0.606
\end{tabular}}
\end{center}
\label{tab:mainresults}
\end{table*}

\subsection{Performance comparison}
Table \ref{tab:mainresults} shows the performance of our model on each of the sessions compared to a number of baselines. Bolded values indicate the best GRU and best non-GRU models for each session (separated by a dashed line). For the SH-GCN model trained with ordinal loss, parentheses indicate the performance of the analogous model which penalized based on the $L_2$ norm between the adjacency matrix and the empirical correlation matrix. We find that the GRU-based models outperform the non-GRU based models - often by a large margin. In addition, non-GRU models which utilize ordinal loss typically outperform those minimizing cross-entropy. Most OR-based SH-GCN models preferred three to five spike history features, affirming the importance of past information in predicting stimulus intensity. This is further illustrated by improvement in performance obtained using the GRU-GCN.

\begin{figure*}[t!]
    \centering
    \begin{subfigure}[t]{0.42\textwidth}
        \centering
        \includegraphics[width=\textwidth]{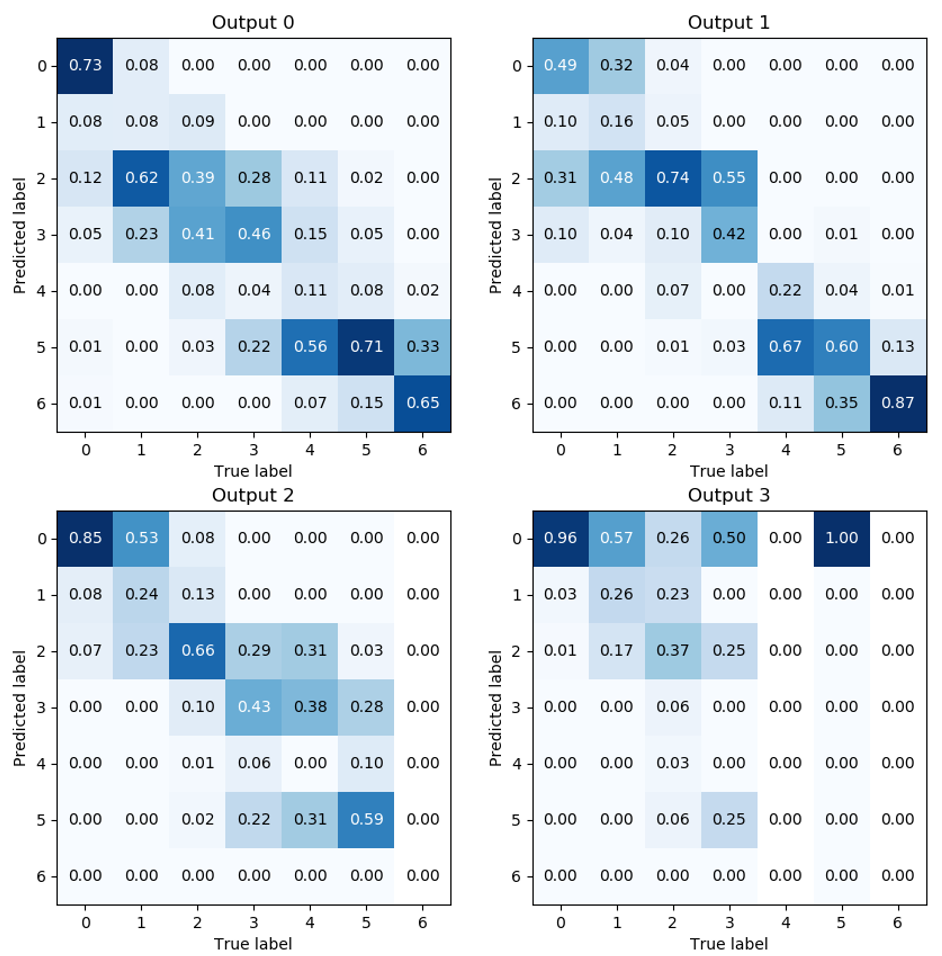}
        \caption{GRU-GCN model trained on OR loss}
    \end{subfigure}
    \hspace{2em}
    \begin{subfigure}[t]{0.42\textwidth}
        \centering
        \includegraphics[width=\textwidth]{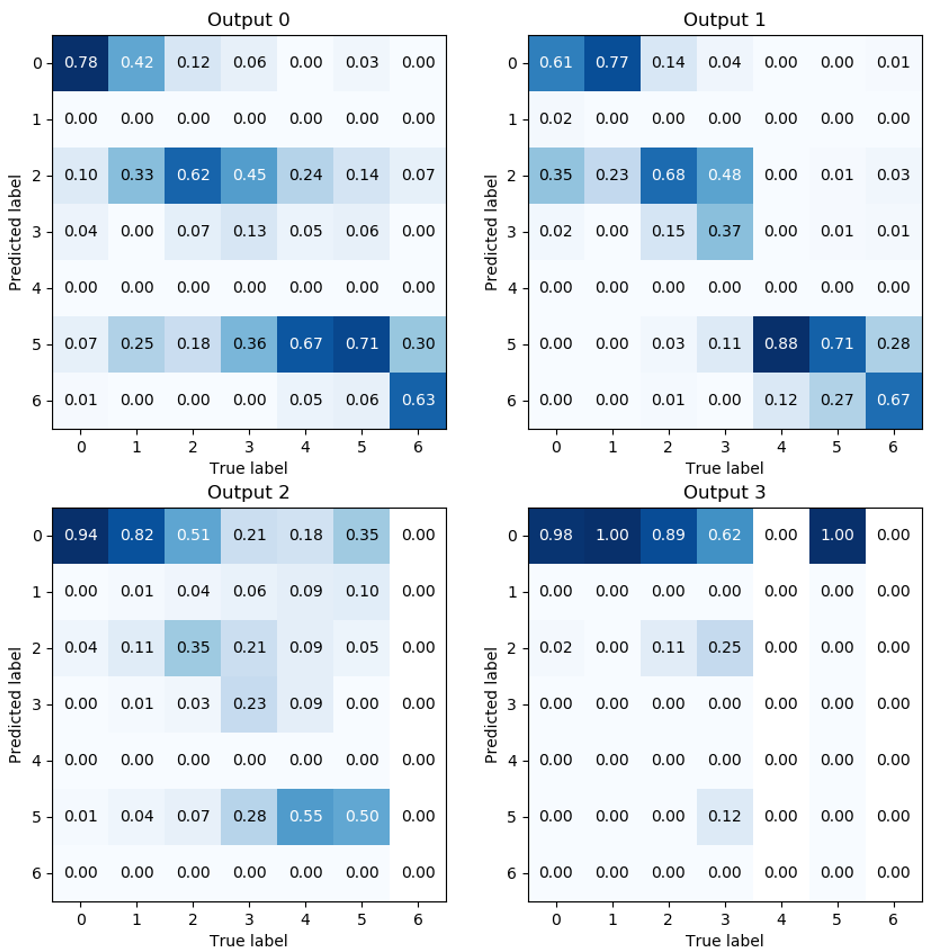}
        \caption{SH-GCN model trained on cross-entropy loss}
    \end{subfigure}
    \caption{Confusion matrices for predictions made by the the top-performing OR-based (GRU-GCN) and cross-entropy-based (SH-GCN) models for session 4. Values in cell $(y, \hat{y})$ correspond to the fraction of points labeled by class $y$ which were predicted to be class $\hat{y}$.}
    \label{fig:conMat}
\end{figure*}

We first considered a na\"{i}ve feedforward network baseline trained with cross-entropy loss. The model consisted of $N=42$ input features corresponding to the number of spikes in the current time bin $t$ for each neuron and $L=3$ hidden layers with 15, 15, and 10 hidden units. To determine whether the graph structure was beneficial to predictive performance, we trained an SH-GCN model, also with $L=3$ hidden layers and with $D=1$ side feature corresponding to the spike count in the $t$\textsuperscript{th} time bin. In three out of seven sessions, the SH-GCN model outperformed the feedforward. We attribute the relatively good performance of the feedforward model to the larger number of network parameters. Specifically, the feedforward network contains separate weight matrices for each hidden layer. On the other hand, the adjacency matrix for the SH-GCN is shared across layers. Moreover, the layer-specific SH-GCN weight matrix reduces to a scalar since only one side feature is employed. The feedforward model, however, does not provide the interpretability offered by the SH-GCN model because it lacks an adjacency matrix.

We next compared the SH-GCN model fit with cross-entropy to SH-GCN fit with OR loss. We did an architecture search over number of hidden layers $L \in \{1, 3\}$ and number of spike history features $D \in \{1, 3, 5\}$. The performance of the best SH-GCN configuration is shown in Table \ref{tab:mainresults}. Enforcing an OR loss resulted in an improvement in $\bar{F}_{OR}$ and $\bar{F}_1$ in a majority of sessions. The OR-based model comes closer to accurately predicting the direction towards the IR source driving the observed neuronal activity. This is because the OR-based model biases the predictions of the stimulus to be closer to the ground truth labels while the cross-entropy model has more variance in its predictions. An improved $\bar{F}_{OR}$ is especially desirable given our objective of decoding the neural activity associated with a particular sensory experience.

One drawback of the SH-GCN model is that the number of parameters grows quadratically with the number of side features. Specifically, if we incorporate $D$ spike history features, we need a $D \times D$ matrix $\mathbf{W}$ to perform a linear map. This motivates the use of a GRU to keep track of the hidden state. The GRU parameters scale linearly with the number of input features, and quadratically with the number of outputs. In our case, we restrict the number of output features (which matches the dimension of the GRU hidden state) to be 10. This is a reasonable compromise between the number of inputs $N=42$ and the number of model outputs (four scalars for ordinal regression on each output). We compared the performance of a baseline GRU model with 10 hidden layers with that of the best SH-GCN model trained with OR loss. Incorporating a hidden state was found to generally improve predictive performance; however, SH-GCN models that preferred more than one previous spike history feature tended to perform just as well.

Finally, we evaluated the performance of the GRU-GCN model, which not only maintains a hidden state but also infers connectivity using a GCN. We trained the model with $L \in \{1, 3\}$ one and three hidden layers and found that in all but one session, the GRU-GCN outperforms the GRU model. In addition, in 5 out of the 7 sessions, the GRU-GCN model outperforms all other models. The hidden state maintains information not only about spiking history but also about past stimulus information. As shown in Figure \ref{fig:conMat}, predictions made by the GRU-GCN model result in confusion matrices whose elements are more concentrated along the diagonal, indicating fewer gross misclassifications.

\begin{figure}[t]
    \centering
    \includegraphics[width=0.6\textwidth]{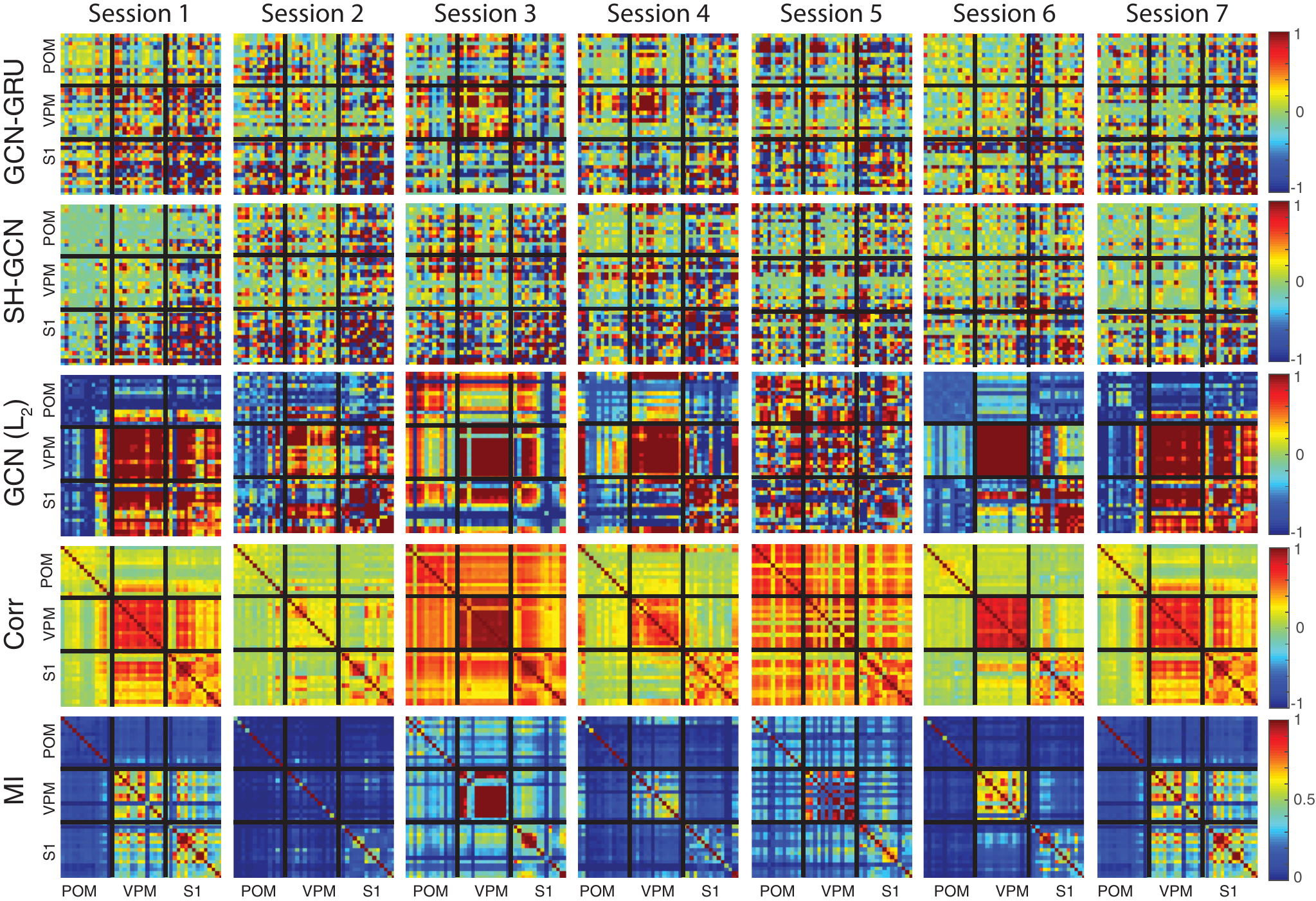}
    \caption{Comparison of the adjacency matrices inferred from our GCN models to other baseline functional connectivity methods. Grid overlay on each matrix delineates the brain area (S1, VPM, or POM) correspondence for each node. Top row shows the adjacency matrix derived for each session (GRU-GCN). The second and third rows show the adjacency matrices derived using the SH-GCN model without and with $L_2$ regularization, respectively. Fourth row shows the correlation coefficients (Corr), and the bottom row shows the mutual information (MI) between each unit.}
    \label{fig:adjCorrMIComp}
\end{figure}

\subsection{Effective Connectivity analysis}
As described previously, we hypothesize that information is transmitted from S1 to VPM and POM during the sensory-augmented navigation task. The MI and correlation matrices for each session show highly correlated activity for units within VPM (Figure \ref{fig:adjCorrMIComp}). This is further captured by the SH-GCN model under $L_2$ regularization constraining the adjacency matrix approximately to correlation coefficients. Importantly, as shown in Table \ref{tab:mainresults}, $L_2$ regularization always decreases $\bar{F}_{OR}$. This suggests that correlation between units is not the best predictor of stimulation frequency. The adjacency matrices for the SH-GCN without $L_2$ regularization and the GRU-GCN models for all sessions show two important features: 1) correlative activity between VPM neurons becomes de-emphasized and 2) networks consisting of S1 units with other neurons become more pronounced. This appears to support our hypothesis that the adjacency matrix inferred by the GCN models will emphasize the networks involved in sensory processing within the augmented navigation task.

In the endogenous circuitry, VPM serves a central role in facilitating processing of whisker deflections. Given that the rat continues to utilize its whisker system while navigating the chamber, the correlated activity within VPM likely encodes this. However, as evidenced by the results of the GCN-based models, these correlations prove uninformative with regards to predicting the stimulus delivered via the sensory prosthesis. The strong edges connecting nodes in VPM and POM to S1 suggest that stimulation delivered to S1 by the prosthesis is being routed to VPM and POM. The correlation matrix provides some sense of connectivity between units; however, statistical measures of significant coactivations between units do not indicate the network created by those units are necessarily involved in the processing of an external stimulus delivered through a sensory prosthesis. Here we are interested in the adjacency matrix inferred by the GCN model because it indicates not only units that have statistical correlations but specifically shows which units show significant correlations in the context of a sensory stimulus. This reveals underlying networks involved in processing the information being projected via the prosthesis.

\section{Conclusions}

In conclusion, we find that GCN-based models perform well in predicting stimulation frequency from neuronal activity when a rat performs a sensory augmented navigation task. Additionally, these models are able to infer effective connectivity between S1, VPM, and POM regions of the brain, helping to explain how neuronal circuitry might encode an artificial sensory modality provided through a prosthesis. Our approach indicates that the sensory circuits for processing the prosthetic sense map onto the natural sensory circuits.

\section{Acknowledgements}

We thank Miguel Nicolelis and Eric Thomson for establishing the IR prosthetic system and behavioral protocol which we used to collect the data presented here and Gautam Nayar for helping conduct experiments. We also thank Lawrence Carin, Hongteng Xu, Nikhil Mehta, Serge Assaad, and Dhanasekar Sundararaman for feedback and helpful discussions.

\bibliographystyle{unsrt}
\bibliography{neurips_2020}

\end{document}